 \documentclass[cameraready]{Interspeech}

\title{Listening Between the Lines: Joint Learning of ASR Embeddings and LLM-Augmented Linguistics for Dementia Detection}

\author[affiliation={1}, orcid=0009-0003-6532-5068, equalcontribution]{Olivier Jiyoun}{Jung}
\author[affiliation={2}, orcid=0009-0003-3544-8040, equalcontribution]{Jonghyeon}{Park}
\author[affiliation={2}, orcid=0009-0007-7683-1117,correspondingauthor]{Myungwoo}{Oh}

\address{
    $^1$ Division of Communication and Media, Ewha Womans University, South Korea \\
    $^2$ NAVER Cloud, South Korea
}

\email{olivierjiyounjung@gmail.com, jong-hyeon.park@navercorp.com, myungwoo.oh@navercorp.com}


\usepackage{comment}

\begin{document}

\maketitle

\begin{abstract}
    
Early detection of dementia through speech analysis offers a non-invasive screening alternative, but capturing both acoustic and linguistic biomarkers remains challenging. We propose a multimodal framework leveraging Whisper for dual-purpose extraction: acoustic representations from encoder outputs and transcripts via automatic speech recognition (ASR). For the acoustic pathway, temporal networks with attention pooling aggregate variable-length sequences into fixed-dimensional embeddings. For the linguistic pathway, we prompt a large language model (LLM) to extract interpretable features spanning lexical diversity, syntactic complexity, semantic coherence, and discourse patterns. A gated fusion network integrates both modalities. On ADReSS and ADReSSo, our method achieves F1-scores of 89.47\% and 90.14\%, demonstrating effective integration of acoustic and LLM-augmented linguistic features. Ablation shows that multimodal fusion consistently outperforms either modality alone.
\end{abstract}

\section{Introduction}

What a patient says and how they say it reflect different but complementary signs of cognitive decline. Yet most detection systems focus on only one of these dimensions. That limitation matters. Dementia affects more than 55 million people worldwide, and Alzheimer's disease (AD) accounts for 60–70\% of cases~\cite{who2023dementia}. Current diagnostic approaches often rely on expensive neuroimaging or invasive biomarker tests, which can limit early and widespread screening.

Speech-based analysis offers a promising non-invasive alternative. Language production is particularly sensitive to the subtle cognitive changes seen in early-stage AD~\cite{mueller2018connected}. To support progress in this area, the ADReSS and ADReSSo challenges~\cite{luz2020adress, luz2021adresso} introduced standardized benchmarks comparing AD patients with cognitively normal (CN) controls based on picture description tasks, allowing researchers to compare detection methods under consistent conditions.

Prior approaches have employed both acoustic and linguistic features for AD detection. Acoustic methods capture paralinguistic markers such as pause patterns and speech rate that differentiate AD from CN speakers~\cite{konig2015automatic}, while linguistic approaches analyze lexical diversity, syntactic complexity, and semantic content~\cite{fraser2016linguistic, ahmed2013connected}. A prominent paradigm is information unit (IU) analysis, which annotates transcripts with predefined content elements from the Cookie Theft picture~\cite{croisile1996comparative}. However, IU-based methods rely on manually constructed coding schemes developed decades ago, which may fail to capture the full range of descriptive content or discourse-level organizational patterns.

The emergence of large language models (LLMs) has opened new possibilities for clinical speech analysis. Recent studies have applied bidirectional encoder representations from transformers (BERT)-based classification~\cite{balagopalan2020bert}, generative pre-trained transformer (GPT)-series models for transcript analysis~\cite{agbavor2022predicting}, and chain-of-thought reasoning for interpretable predictions~\cite{park2025reasoning}. Despite these advances, most LLM-based approaches either treat embeddings as black-box features or simply match against existing IU frameworks rather than leveraging LLMs to derive comprehensive, interpretable feature sets. Furthermore, acoustic and linguistic modalities are often processed in isolation, neglecting their complementary nature for dementia detection.

We propose a multimodal framework that addresses these limitations by integrating acoustic and linguistic representations. We leverage Whisper~\cite{radford2023whisper} as a dual-purpose module, extracting both acoustic representations from its encoder outputs and transcripts via automatic speech recognition (ASR). For the acoustic pathway, temporal networks with attention pooling~\cite{bahdanau2015attention} aggregate variable-length sequences into fixed-dimensional vectors. For the linguistic pathway, we prompt an LLM to extract 46 interpretable features spanning lexical diversity, syntactic complexity, semantic coherence, and discourse-level cognitive indicators. Feature selection yields an optimized 29-feature subset. A gated fusion network~\cite{arevalo2017gated} dynamically integrates both modalities, allowing the model to adaptively weight acoustic and linguistic contributions. Notably, feature subsets including statistically non-significant features outperform significance-filtered subsets, demonstrating the importance of feature interactions in classification. Our contributions are threefold:
\begin{itemize}
    \item A multimodal framework that leverages Whisper~\cite{radford2023whisper} for dual-purpose feature extraction and combines attention-pooled~\cite{bahdanau2015attention} acoustic representations with LLM-augmented linguistic features through gated fusion~\cite{arevalo2017gated}.
    \item A comprehensive set of interpretable linguistic features extracted via LLM prompting, capturing multiple dimensions of discourse impairment.
    \item Empirical evidence that statistically non-significant features contribute through feature interactions, with our method achieving an F1-score of 90.14\% on ADReSSo for AD versus CN classification.
\end{itemize}

\section{Methods}

\begin{figure*}[t]
  \centering
  \includegraphics[width=\textwidth]{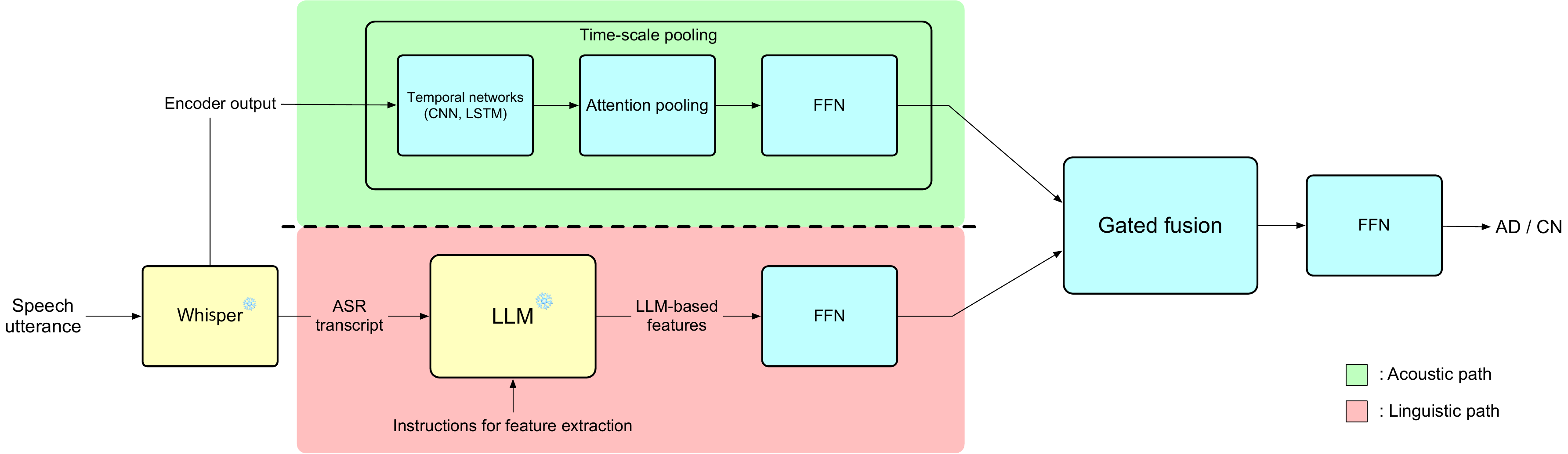}
  \caption{Overview of the proposed multimodal framework. The acoustic pathway extracts attention-pooled~\cite{bahdanau2015attention} representations from Whisper~\cite{radford2023whisper} encoder outputs via temporal networks. The linguistic pathway derives interpretable features through LLM-based sentence classification. A gated fusion network~\cite{arevalo2017gated} integrates both modalities for AD/CN classification.}
  \label{fig:overall_diagram}
\end{figure*}

\begin{table*}[t]
\centering
\caption{Hierarchical topic taxonomy for Cookie Theft picture description}
\label{tab:taxonomy}
\footnotesize
\begin{tabular}{@{}cl@{\qquad}cl@{}}
\toprule
\textbf{Cluster} & \textbf{Attentional Zone} & \textbf{Cluster} & \textbf{Attentional Zone} \\
\midrule
C1 & Boy \& Cabinet Action       & C5 & Counter Objects \\
C2 & Girl Participation          & C6 & Window \& Exterior \\
C3 & Mother \& Domestic Activity & C7 & Room Layout \& Setting \\
C4 & Water Overflow Event        & C8 & Meta-discourse \\
\bottomrule
\end{tabular}
\end{table*}

\subsection{Overview}

Figure~\ref{fig:overall_diagram} illustrates our multimodal framework. Given a speech recording, we use Whisper~\cite{radford2023whisper} large-v3 for dual-purpose feature extraction: encoder outputs serve as acoustic representations, while the decoder produces transcripts for linguistic analysis.

The framework comprises two parallel pathways. The acoustic pathway processes Whisper encoder outputs through temporal networks followed by attention pooling~\cite{bahdanau2015attention} to obtain fixed-dimensional speech representations. The linguistic pathway prompts GPT-5.2~\cite{openai2025gpt5} with ASR transcripts and structured instructions to extract interpretable features spanning lexical, syntactic, semantic, and discourse dimensions. A gated fusion network~\cite{arevalo2017gated} integrates both modalities, learning to weight their relative contributions for classification. We describe each component in  detail below.

\subsection{Acoustic Feature Extraction}

We extract frame-level representations from the Whisper~\cite{radford2023whisper} large-v3 encoder. These representations capture spectral characteristics, temporal dynamics, and prosodic patterns relevant to cognitive assessment.

To aggregate variable-length frame sequences into fixed-dimensional representations, we employ temporal networks such as a convolutional neural network (CNN)~\cite{lecun1998gradient} or a bidirectional long short-term memory (LSTM)~\cite{hochreiter1997lstm,schuster1997bidirectional} followed by attention pooling~\cite{bahdanau2015attention}. Rather than using only the final hidden state, we compute attention weights over all time steps:

\begin{equation}
\alpha_t = \frac{\exp(\mathbf{w}^\top \tanh(\mathbf{W} h_t))}{\sum_{t'} \exp(\mathbf{w}^\top \tanh(\mathbf{W} h_{t'}))}
\end{equation}
\begin{equation}
\mathbf{c} = \sum_t \alpha_t h_t
\end{equation}

\noindent where $h_t$ denotes the concatenated bidirectional hidden state at time step $t$, and $\mathbf{W}$, $\mathbf{w}$ are learnable parameters. This mechanism enables the model to focus on diagnostically relevant temporal segments. The pooled representation passes through a feed-forward network with LayerNorm~\cite{ba2016layernorm} to produce the final acoustic feature vector.

\subsection{LLM-Augmented Linguistic Features}

\subsubsection{Hierarchical Topic Taxonomy Construction}

Picture description analysis requires a coding scheme that captures the semantic structure of the stimulus. Prior approaches rely on manually constructed IU~\cite{croisile1996comparative,ahmed2013connected} or keyword lists, which may incompletely capture descriptive content and lack systematic organization.

We constructed a hierarchical topic taxonomy\footnote{The details are provided in https://github.com/vivivic/is26dementia.} using GPT-5.2~\cite{openai2025gpt5}. The Cookie Theft picture from the Boston Diagnostic Aphasia Examination (BDAE)~\cite{goodglass1983boston} was provided with a prompt requesting topic categories organized by spatial proximity, agent-centered grouping, and thematic coherence. The model defined clusters as ``attentional zones''---regions or themes that speakers naturally group together.

The resulting taxonomy comprises eight clusters, as detailed in Table~\ref{tab:taxonomy}. Clusters C1–C7 capture content related to principal agents, salient events, objects, and spatial context, whereas Cluster C8 encompasses meta-discourse, including filled pauses, uncertainty markers, and self-corrections.

\subsubsection{Unified Sentence-Level Classification}

Transcripts annotated with the CHAT coding system~\cite{macwhinney2000childes} are preprocessed to isolate participant utterances by removing disfluency annotations and markup tags. Rather than using separate classification passes, we designed a unified prompt that extracts all annotations in a single inference call, as illustrated in Figure~\ref{fig:prompt}, ensuring consistency while improving computational efficiency.

The classification schema covers five dimensions:

\noindent\textbf{Topic Classification} assigns each sentence to its primary cluster (C1--C8) and specific topic.

\noindent\textbf{Classification Confidence} captures certainty about topic assignment, utterance completeness, and task relevance.

\noindent\textbf{Language Quality} ratings, motivated by the 7-point scales for grammatical form and fluency in the BDAE Rating Scale Profile of Speech Characteristics~\cite{goodglass1983boston}, highlight a dissociation in AD speech where syntactic structure is preserved while verbal fluency is significantly reduced.

\noindent\textbf{Content Integration} characterizes whether descriptions integrate multiple clusters or maintain single focus, specifying relationship types (causal, temporal, spatial, contrastive).

\noindent\textbf{Semantic Distance} serves as a computational index to quantify coherence, while manual evaluation relies on a three-point global coherence scale adapted from the original five-point version~\cite{glosser1991, vanleer1999}.

\begin{figure}[t]
\centering
\fbox{\parbox{0.93\columnwidth}{
\footnotesize
\setlength{\baselineskip}{0.95\baselineskip}
\scriptsize
\texttt{[TASK]}\\
You are analyzing a Cookie Theft picture description transcript.
For EACH sentence, provide ALL analyses in a single JSON response.\\[0.1em]

\texttt{[CLUSTERS]}\\
C1: boy\_on\_stool, reaching\_up, cookie\_jar, ...\\
C2: girl\_reaching, asking\_for\_cookie, ...\\
\hspace{1em}$\vdots$\\
C8: meta\_filler, uncertainty, self\_correction, ...\\[0.1em]

\texttt{[DIMENSIONS]}\\
1. \textbf{Topic}: cluster (C1--C8), specific topic\\
2. \textbf{Confidence}: high $|$ medium $|$ low\\
3. \textbf{Language Quality}: grammaticality, fluency (1--7)\\
4. \textbf{Content Integration}: integrated $|$ single\_focus\\
5. \textbf{Semantic Distance}: 1 (related) $|$ 2 $|$ 3 (abrupt)\\[0.1em]

\texttt{[OUTPUT]} Return a valid JSON array.
}}
\caption{Unified prompt for multi-dimensional sentence annotation.}
\label{fig:prompt}
\end{figure}

\subsubsection{Feature Computation}

From sentence-level classifications, we compute 46 speaker-level features across six categories, as summarized in Table~\ref{tab:features}. We then select an optimized subset of 29 features based on classification performance rather than statistical significance alone; Section~\ref{sec:ablation} details the selection rationale.

\begin{table}[t]
\centering
\caption{Complete feature inventory (46 features)}
\label{tab:features}
\footnotesize
\setlength{\tabcolsep}{1pt}
\begin{tabular}{@{}lc@{}}
\toprule
\textbf{Category} & \textbf{Count} \\
\midrule
Discourse Diversity (entropy, coverage, C1--C8 ratios) & 13 \\
Discourse Flow (transition, revisit, jump rates) & 5 \\
Language Quality (grammaticality, fluency statistics) & 8 \\
Content Integration (ratios, relationship counts) & 9 \\
Classification Confidence (certainty, completeness) & 10 \\
Meta (sentence count) & 1 \\
\midrule
\textbf{Total} & \textbf{46} \\
\bottomrule
\end{tabular}
\end{table}

\subsection{Gated Multimodal Fusion}

To integrate acoustic and linguistic representations adaptively, we employ a gated fusion mechanism~\cite{arevalo2017gated} that learns the relative importance of each modality. Given acoustic representation $\mathbf{s}$ and linguistic representation $\mathbf{f}$, the fusion gate is computed as:

\begin{equation}
\mathbf{g} = \sigma(\mathbf{W}_g [\mathbf{s}; \mathbf{f}] + \mathbf{b}_g)
\end{equation}

\noindent where $\sigma$ denotes the sigmoid function and $[\cdot;\cdot]$ represents concatenation. The fused representation is:

\begin{equation}
\mathbf{z} = \mathbf{g} \odot \mathbf{s} + (1 - \mathbf{g}) \odot \mathbf{f}
\end{equation}

\noindent where $\odot$ denotes element-wise multiplication. This gating mechanism enables the model to dynamically weight each modality based on its informativeness for each sample. The fused representation passes through a feed-forward classifier with LayerNorm~\cite{ba2016layernorm} to produce AD/CN predictions.

\section{Experimental Settings}

\subsection{Dataset}

We evaluate on two benchmark datasets from the ADReSS challenge series~\cite{luz2020adress,luz2021adresso}, both derived from DementiaBank's Pitt Corpus~\cite{becker1994natural}. The corpora comprise audio recordings of participants performing the Cookie Theft picture description task from the BDAE~\cite{goodglass1983boston}. Both datasets provide transcripts annotated using CHAT coding conventions~\cite{macwhinney2000childes} with detailed disfluency markers.

\textbf{ADReSS}~\cite{luz2020adress} contains 156 speakers balanced by age, gender, and diagnosis: 108 for training (54 AD, 54 CN) and 48 for testing (24 AD, 24 CN).

\textbf{ADReSSo}~\cite{luz2021adresso} provides additional speakers without demographic balancing: 166 for training and 71 for testing.

\subsection{Implementation Details}

\textbf{Acoustic pathway.} Whisper~\cite{radford2023whisper} large-v3 encoder outputs yield 1280-dimensional frame-level representations. The bidirectional LSTM~\cite{hochreiter1997lstm,schuster1997bidirectional} comprises two layers with hidden dimension 128, producing 256-dimensional concatenated states. Attention pooling and a subsequent feed-forward network output a 128-dimensional acoustic vector.

\textbf{Linguistic pathway.} We use GPT-5.2~\cite{openai2025gpt5} to extract 29 features from transcripts. These pass through a feed-forward network (hidden dimension 32) with LayerNorm~\cite{ba2016layernorm}, yielding a 128-dimensional linguistic vector.

\textbf{Fusion and classification.} The gated fusion network~\cite{arevalo2017gated} operates on the concatenated 256-dimensional input to produce 128-dimensional fused representations. The classifier consists of two feed-forward layers (hidden dimension 64) with LayerNorm.

\textbf{Training.} We optimize using AdamW~\cite{loshchilov2019decoupled} with learning rate $2 \times 10^{-5}$ and batch size 64. Training employs early stopping with patience of 30 epochs based on validation F1-score. For evaluation, we aggregate segment-level predictions to speaker-level via majority voting.

\section{Results}

\subsection{Main Results}

Table~\ref{tab:main_results} presents speaker-level classification performance on both benchmark datasets.

\begin{table}[t]
\centering
\caption{Speaker-level classification performance on ADReSS and ADReSSo test sets. P: Precision, R: Recall.}
\label{tab:main_results}
\small
\setlength{\tabcolsep}{5pt}
\begin{tabular}{@{}lcccccc@{}}
\toprule
\textbf{Dataset} & \textbf{Acc} & \textbf{F1} & \textbf{P\textsubscript{AD}} & \textbf{R\textsubscript{AD}} & \textbf{P\textsubscript{CN}} & \textbf{R\textsubscript{CN}} \\
\midrule
ADReSS   & 89.58 & 89.47 & 100.0 & 79.17 & 82.76 & 100.0 \\
ADReSSo  & 90.14 & 90.14 & 88.89 & 91.43 & 91.43 & 88.89 \\
\bottomrule
\end{tabular}
\end{table}

Our method achieves strong performance on both benchmarks. On ADReSS, the model exhibits high precision for AD detection (100\%) but lower recall (79.17\%), indicating a conservative classification threshold that minimizes false positives. On ADReSSo, precision and recall are balanced across both classes, suggesting robust generalization to the unbalanced demographic distribution.
Clinically, several features with large effect sizes are interpretable as markers of cognitive decline: elevated \texttt{C8\_ratio} (meta-discourse such as filled pauses, uncertainty, and self-corrections) reflects word-finding difficulty in AD~\cite{glosser1991}, while reduced \texttt{cluster\_coverage\_ratio} indicates that fewer attentional zones are described, consistent with impaired discourse planning.

\subsection{Comparison with Prior Work}

To contextualize these results, Table~\ref{tab:comparison} compares our method with recent acoustic, linguistic, and multimodal approaches on both benchmarks.

\begin{table}[t]
\centering
\caption{Comparison with prior work on ADReSS and ADReSSo official test sets (F1-score, \%). A: Acoustic, L: Linguistic, M: Multimodal.}
\label{tab:comparison}
\small
\begin{tabular}{@{}llcc@{}}
\toprule
\textbf{Method} & \textbf{Type} & \textbf{ADReSS} & \textbf{ADReSSo} \\
\midrule
Luz et al. (2020)~\cite{luz2020adress} & M & 75.00 & -- \\
Luz et al. (2021)~\cite{luz2021adresso} & M & -- & 78.87 \\
Zhu et al. (2021)~\cite{zhu2021wavbert} & M & -- & 83.02 \\
Ilias et al. (2022)~\cite{ilias2022multimodal} & M & 85.48 & -- \\
Li \& Zhang (2024)~\cite{li2024whisper} & A & -- & 84.51 \\
Park et al. (2025)~\cite{park2025reasoning} & L & 87.50 & -- \\
\midrule
Ours (Linguistic only) & L & -- & 76.06 \\
Ours (Acoustic only)   & A & -- & 83.08 \\
Ours (Multimodal) & M & \textbf{89.47} & \textbf{90.14} \\
\bottomrule
\end{tabular}
\end{table}

Our multimodal approach outperforms all compared methods, yielding relative F1-score improvements of 19.3\% and 14.3\% over the official challenge baselines on ADReSS and ADReSSo, respectively. These gains derive from two factors: (1) using LLM-based reasoning to construct a hierarchical topic taxonomy rather than relying on predefined IUs, and (2) combining Whisper~\cite{radford2023whisper}-based acoustic representations with LLM-augmented linguistic features through gated fusion~\cite{arevalo2017gated}.

Acoustic features alone substantially outperform linguistic features (83.08\% vs. 76.06\% F1-score on ADReSSo), indicating that paralinguistic markers carry strong discriminative signal. However, multimodal fusion yields gains of 7.1 and 14.1 percentage points over the respective unimodal baselines, supporting the view that acoustic and linguistic biomarkers capture complementary aspects of cognitive decline.

\subsection{Ablation Studies}
\label{sec:ablation}

\subsubsection{Feature Selection}

We evaluate three feature configurations to examine the relationship between statistical significance and classification performance, as shown in Table~\ref{tab:feature_ablation}.

\begin{table}[t]
\centering
\caption{Feature selection ablation on ADReSSo. Statistical significance determined by independent t-test ($\alpha = 0.05$).}
\label{tab:feature_ablation}
\small
\setlength{\tabcolsep}{4pt}
\begin{tabular}{@{}lccccc@{}}
\toprule
\textbf{Configuration} & \textbf{Feat.} & \textbf{Sig.} & \textbf{$|\bar{d}|$} & \textbf{F1} & \textbf{Acc} \\
\midrule
All features     & 46 & 26 (56.5\%) & 0.393 & 88.69 & 88.73 \\
Significant only & 26 & 26 (100\%)  & 0.568 & 78.82 & 78.87 \\
Optimized        & 29 & 13 (44.8\%) & 0.375 & \textbf{90.14} & \textbf{90.14} \\
\bottomrule
\end{tabular}
\end{table}

The optimized 29-feature subset was identified through intensive experimentation, guided by domain considerations: retaining complete cluster distribution ratios (C1--C8) to preserve attentional profiles, selecting variability measures over central tendencies based on the hypothesis that inconsistency distinguishes groups, and including discourse dynamics features to capture sequential organization.

Restricting features to only those with significant group differences ($p < 0.05$) yields the highest average effect size but substantially degrades performance. The optimized subset—where only 44.8\% of features show individual significance—achieves the best results. This demonstrates that features lacking individual discriminative power contribute through feature interactions, highlighting the limitation of univariate selection for multivariate classification.

\subsubsection{Temporal Architecture}

Table~\ref{tab:arch_ablation} compares temporal architectures for acoustic feature processing. LSTM~\cite{hochreiter1997lstm,schuster1997bidirectional} outperforms CNN~\cite{lecun1998gradient} on ADReSSo while achieving equivalent performance on ADReSS. This likely reflects LSTM's capacity to model long-range dependencies: ADReSSo provides only timestamps, requiring longer variable-length sequences, whereas ADReSS supplies shorter pre-segmented chunks that reduce the benefit of sequential modeling.

\begin{table}[t]
\centering
\caption{Temporal architecture ablation on ADReSS and ADReSSo. Performance reported as F1-score (\%).}
\label{tab:arch_ablation}
\small
\begin{tabular}{@{}lcc@{}}
\toprule
\textbf{Architecture} & \textbf{ADReSS} & \textbf{ADReSSo} \\
\midrule
CNN  & 89.47 & 88.73 \\
LSTM & 89.47 & \textbf{90.14} \\
\bottomrule
\end{tabular}
\end{table}

\section{Conclusion}

We presented a multimodal framework for dementia detection that integrates Whisper~\cite{radford2023whisper}-based acoustic representations with LLM-augmented linguistic features through gated fusion~\cite{arevalo2017gated}. Our key contribution is leveraging LLM reasoning to automatically construct a hierarchical topic taxonomy for picture description analysis, eliminating dependence on manually defined IUs while enabling extraction of interpretable features that capture discourse-level patterns of cognitive decline.

Our method achieves outstanding performance on both ADReSS and ADReSSo benchmarks. Ablation studies reveal that statistically non-significant features contribute through feature interactions, and that multimodal fusion substantially outperforms either modality alone—supporting the complementary nature of acoustic and linguistic biomarkers.

Several limitations remain. The LLM-based feature extraction requires API access, posing challenges for deployment in resource-constrained clinical settings. Additionally, our evaluation is restricted to English-language Cookie Theft descriptions; generalization to other languages and elicitation tasks remains unexplored. Future work will investigate lightweight local models for on-device inference, extend evaluation to multilingual benchmarks, and incorporate longitudinal analysis to track disease progression.

\section{Generative AI Use Disclosure}

We used generative AI for extracting the LLM-augmented linguistic features described in Section 2.3. GPT-5.2~\cite{openai2025gpt5} was consistently used for feature extraction, and the instructions for extraction are provided in https://github.com/vivivic/is26dementia.


\bibliographystyle{IEEEtran}
\bibliography{mybib}

@misc{who2023dementia,
  title={Dementia},
  author={{World Health Organization}},
  year={2023},
  howpublished={\url{https://www.who.int/news-room/fact-sheets/detail/dementia}},
  note={Accessed: 2025}
}

@article{mueller2018connected,
  title={Connected Speech and Language in Mild Cognitive Impairment and Alzheimer's Disease: A Review of Picture Description Tasks},
  author={Mueller, Kimberly D and Hermann, Bruce and Mecollari, Jonilda and Turkstra, Lyn S},
  journal={Journal of Clinical and Experimental Neuropsychology},
  volume={40},
  number={9},
  pages={917--939},
  year={2018},
  publisher={Taylor \& Francis}
}

@inproceedings{luz2020adress,
  title={Alzheimer's Dementia Recognition through Spontaneous Speech: The {ADReSS} Challenge},
  author={Luz, Saturnino and Haider, Fasih and de la Fuente, Sofia and Fromm, Davida and MacWhinney, Brian},
  booktitle={Proceedings of INTERSPEECH},
  pages={2172--2176},
  year={2020},
  doi={10.21437/Interspeech.2020-2571}
}

@inproceedings{luz2021adresso,
  title={Detecting Cognitive Decline Using Speech Only: The {ADReSSo} Challenge},
  author={Luz, Saturnino and Haider, Fasih and de la Fuente, Sofia and Fromm, Davida and MacWhinney, Brian},
  booktitle={Proceedings of INTERSPEECH},
  pages={3780--3784},
  year={2021},
  doi={10.21437/Interspeech.2021-1220}
}

@article{konig2015automatic,
  title={Automatic Speech Analysis for the Assessment of Patients with Predementia and {Alzheimer's} Disease},
  author={K{\"o}nig, Alexandra and Satt, Aharon and Sorin, Alexander and Hoory, Ron and Toledo-Ronen, Orith and Derreumaux, Alexandre and Manera, Valeria and Verhey, Frans and Aalten, Pauline and Robert, Philippe H and David, Renaud},
  journal={Alzheimer's \& Dementia: Diagnosis, Assessment \& Disease Monitoring},
  volume={1},
  number={1},
  pages={112--124},
  year={2015},
  publisher={Elsevier}
}

@article{fraser2016linguistic,
  title={Linguistic Features Identify Alzheimer's Disease in Narrative Speech},
  author={Fraser, Kathleen C and Meltzer, Jed A and Rudzicz, Frank},
  journal={Journal of Alzheimer's Disease},
  volume={49},
  number={2},
  pages={407--422},
  year={2016},
  publisher={IOS Press}
}

@article{ahmed2013connected,
  title={Connected Speech as a Marker of Disease Progression in Autopsy-Proven {Alzheimer's} Disease},
  author={Ahmed, Samrah and Haigh, Anne-Marie F and de Jager, Celeste A and Garrard, Peter},
  journal={Brain},
  volume={136},
  number={12},
  pages={3727--3737},
  year={2013},
  publisher={Oxford University Press}
}

@article{croisile1996comparative,
  title={Comparative Study of Oral and Written Picture Description in Patients with {Alzheimer's} Disease},
  author={Croisile, Bernard and Ska, Bernadette and Brabant, Marie-Jos{\'e}e and Duchene, Annick and Lepage, Yves and Aimard, G{\'e}rard and Trillet, Marc},
  journal={Brain and Language},
  volume={53},
  number={1},
  pages={1--19},
  year={1996},
  publisher={Elsevier}
}

@inproceedings{balagopalan2020bert,
  title={To {BERT} or Not To {BERT}: Comparing Speech and Language-based Approaches for {Alzheimer's} Disease Detection},
  author={Balagopalan, Aparna and Eyre, Benjamin and Rudzicz, Frank and Novikova, Jekaterina},
  booktitle={Proceedings of INTERSPEECH},
  pages={2167--2171},
  year={2020},
  doi={10.21437/Interspeech.2020-2557}
}

@article{agbavor2022predicting,
  title={Predicting Dementia from Spontaneous Speech Using Large Language Models},
  author={Agbavor, Felix and Liang, Hualou},
  journal={PLOS Digital Health},
  volume={1},
  number={12},
  pages={e0000168},
  year={2022},
  publisher={Public Library of Science}
}

@inproceedings{park2025reasoning,
  title={Reasoning-Based Approach with Chain-of-Thought for {Alzheimer's} Detection Using Speech and Large Language Models},
  author={Park, Chanwoo and Choi, Anna Seo Gyeong and Cho, Sunghye and Kim, Chanwoo},
  booktitle={Proceedings of INTERSPEECH},
  year={2025}
}

@inproceedings{radford2023whisper,
  title={Robust Speech Recognition via Large-Scale Weak Supervision},
  author={Radford, Alec and Kim, Jong Wook and Xu, Tao and Brockman, Greg and McLeavey, Christine and Sutskever, Ilya},
  booktitle={International Conference on Machine Learning (ICML)},
  pages={28492--28518},
  year={2023},
  organization={PMLR}
}

@misc{openai2025gpt5,
  title={GPT-5.2 System Card},
  author={OpenAI},
  year={2025},
  url={https://cdn.openai.com/pdf/3a4153c8-c748-4b71-8e31-aecbde944f8d/oai_5_2_system-card.pdf}
}

@inproceedings{bahdanau2015attention,
  title={Neural Machine Translation by Jointly Learning to Align and Translate},
  author={Bahdanau, Dzmitry and Cho, Kyunghyun and Bengio, Yoshua},
  booktitle={International Conference on Learning Representations (ICLR)},
  year={2015}
}

@article{lecun1998gradient,
  title={Gradient-Based Learning Applied to Document Recognition},
  author={LeCun, Yann and Bottou, L{\'e}on and Bengio, Yoshua and Haffner, Patrick},
  journal={Proceedings of the IEEE},
  volume={86},
  number={11},
  pages={2278--2324},
  year={1998}
}

@article{hochreiter1997lstm,
  title={Long Short-Term Memory},
  author={Hochreiter, Sepp and Schmidhuber, J{\"u}rgen},
  journal={Neural Computation},
  volume={9},
  number={8},
  pages={1735--1780},
  year={1997},
  publisher={MIT Press}
}

@article{schuster1997bidirectional,
  title={Bidirectional Recurrent Neural Networks},
  author={Schuster, Mike and Paliwal, Kuldip K},
  journal={IEEE Transactions on Signal Processing},
  volume={45},
  number={11},
  pages={2673--2681},
  year={1997},
  publisher={IEEE}
}

@article{ba2016layernorm,
  title={Layer Normalization},
  author={Ba, Jimmy Lei and Kiros, Jamie Ryan and Hinton, Geoffrey E},
  journal={arXiv preprint arXiv:1607.06450},
  year={2016}
}

@book{goodglass1983boston,
  title={Boston Diagnostic Aphasia Examination},
  author={Goodglass, Harold and Kaplan, Edith},
  year={1983},
  publisher={Lea \& Febiger},
  address={Philadelphia}
}

@book{macwhinney2000childes,
  title={The {CHILDES} Project: Tools for Analyzing Talk},
  author={MacWhinney, Brian},
  year={2000},
  edition={3rd},
  publisher={Lawrence Erlbaum Associates},
  address={Mahwah, NJ}
}

@article{glosser1991,
  title={Patterns of Discourse Production among Neurological Patients with Fluent Language Disorders},
  author={Glosser, Guila and Deser, Toni},
  journal={Brain and Language},
  volume={40},
  number={1},
  pages={67--88},
  year={1991},
  publisher={Elsevier},
  doi={10.1016/0093-934X(91)90117-J}
}

@article{vanleer1999,
  title={The Effect of Elicitation Task on Discourse Coherence and Cohesion in Adolescents with Brain Injury},
  author={Van Leer, Eva and Turkstra, Lyn},
  journal={Journal of Communication Disorders},
  volume={32},
  number={5},
  pages={327--349},
  year={1999},
  publisher={Elsevier},
  doi={10.1016/S0021-9924(99)00008-8}
}

@inproceedings{arevalo2017gated,
  title={Gated Multimodal Units for Information Fusion},
  author={Arevalo, John and Solorio, Thamar and Montes-y-G{\'o}mez, Manuel and Gonz{\'a}lez, Fabio A},
  booktitle={International Conference on Learning Representations (ICLR) Workshop},
  year={2017}
}

@article{becker1994natural,
  title={The Natural History of {Alzheimer's} Disease: Description of Study Cohort and Accuracy of Diagnosis},
  author={Becker, James T and Boller, Fran{\c{c}}ois and Lopez, Oscar L and Saxton, Judith and McGonigle, Karen L},
  journal={Archives of Neurology},
  volume={51},
  number={6},
  pages={585--594},
  year={1994},
  publisher={American Medical Association}
}

@inproceedings{zhu2021wavbert,
  title={{WavBERT}: Exploiting Semantic and Non-semantic Speech using {Wav2vec} and {BERT} for Dementia Detection},
  author={Zhu, Youxiang and Obyat, Abdelrahman and Liang, Xiaohui and Batsis, John A and Roth, Robert M},
  booktitle={Proceedings of INTERSPEECH},
  pages={3790--3794},
  year={2021},
  doi={10.21437/Interspeech.2021-332}
}

@article{ilias2022multimodal,
  title={A Multimodal Approach for Dementia Detection from Spontaneous Speech with Tensor Fusion Layer},
  author={Ilias, Loukas and Askounis, Dimitris and Psarras, John},
  journal={arXiv preprint arXiv:2211.04368},
  year={2022}
}

@inproceedings{li2024whisper,
  title={Whisper-Based Transfer Learning for {Alzheimer} Disease Classification: Leveraging Speech Segments with Full Transcripts as Prompts},
  author={Li, Jinpeng and Zhang, Wei-Qiang},
  booktitle={IEEE International Conference on Acoustics, Speech and Signal Processing (ICASSP)},
  pages={11211--11215},
  year={2024},
  organization={IEEE},
  doi={10.1109/ICASSP48485.2024.10448004}
}

@inproceedings{loshchilov2019decoupled,
  title={Decoupled Weight Decay Regularization},
  author={Loshchilov, Ilya and Hutter, Frank},
  booktitle={Proceedings of the 7th International Conference on Learning Representations (ICLR)},
  year={2019}
}

\end{document}